\documentclass{article}
\usepackage{arxiv}
\usepackage{hyperref}
\usepackage{url}
\usepackage[utf8]{inputenc}
\usepackage{booktabs}
\usepackage{amsmath}
\usepackage{amsfonts}
\usepackage{amssymb}
\usepackage[inline]{enumitem}
\usepackage{enumitem}
\usepackage{mathbbol}
\usepackage[T1]{fontenc}
\usepackage{mathtools}
\usepackage{multirow}
\usepackage[square,sort,comma,numbers]{natbib}
\usepackage{graphicx}
\usepackage{titlesec}
\usepackage{xcolor}

\newcommand{\ie}{\emph{i.e.}}
\newcommand{\eg}{\emph{e.g.}}

\titleclass{\subsubsubsection}{straight}[\subsection]
\newcounter{subsubsubsection}[subsubsection]
\renewcommand\thesubsubsubsection{\thesubsubsection.\arabic{subsubsubsection}}

\titleformat{\subsubsubsection}
  {\normalfont\normalsize\bfseries}{\thesubsubsubsection}{1em}{}
\titlespacing*{\subsubsubsection}
{0pt}{3.25ex plus 1ex minus .2ex}{1.5ex plus .2ex}

\title{Sociotechnical Implications of Generative Artificial Intelligence for Information Access}

\author{
\href{https://orcid.org/my-orcid?orcid=0000-0002-5270-5550}
{\includegraphics[scale=0.06]{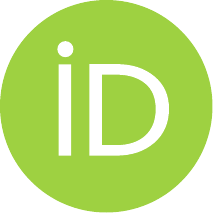}\hspace{1mm}Bhaskar Mitra} \\
Microsoft Research \\
Montréal, Canada \\
\texttt{bmitra@microsoft.com}
\and
Henriette Cramer \\
PaperMoon AI \\
San Francisco, USA \\
\texttt{henriette.cramer@gmail.com}
\and
Olya Gurevich \\
PaperMoon AI \\
San Francisco, USA \\
\texttt{olya.gurevich@gmail.com}
}
\date{\vspace{-6ex}}

\begin{document}
\maketitle

\begin{abstract}
Robust access to trustworthy information is a critical need for society with implications for knowledge production, public health education, and promoting informed citizenry in democratic societies.
Generative AI technologies may enable new ways to access information and improve effectiveness of existing information retrieval systems but we are only starting to understand and grapple with their long-term social implications.
In this chapter, we present an overview of some of the systemic consequences and risks of employing generative AI in the context of information access.
We also provide recommendations for evaluation and mitigation, and discuss challenges for future research.
\end{abstract}
\section{Introduction}
\label{sec:intro}
Robust access to trustworthy information is a critical need for society including implications for knowledge production, public health education, and promoting informed citizenry in democratic societies.
Generative AI technologies such as large language models (LLMs) may enable new ways to access information and improve effectiveness of existing information retrieval (IR) systems.
More efficient basic task execution with the help of LLMs can also enable people to focus on the more challenging aspects of information retrieval related tasks and research.
However, the long-term social implications of deploying these technologies in the context of information access are not yet well-understood.
Existing research has focused on how these models may generate biased and harmful content~\citep{bender2021dangers, abid2021persistent, ferrara2023should, navigli2023biases, kotek2023gender, welbl2021challenges, gehman2020realtoxicityprompts} as well as the environmental costs~\citep{bender2021dangers, patterson2021carbon, bommasani2021opportunities, wu2022sustainable, dodge2022measuring, patterson2022carbon} of developing and deploying these models at scale.
In the context of information access, \citet{shah2022situating} have argued that certain framings of LLMs as ``search engines'' lack the necessary theoretical underpinnings and may constitute as a category error.

In this current work, we present a broader perspective on the sociotechnical implications of generative AI for information access.
Our perspective is informed by existing literature and aims to provide a summary of known challenges viewed through a systemic lens that we hope will serve as a useful resource for future critical research in this area.
We present a summary of these implications next followed by recommendations for evaluation and mitigation later in this chapter.
\section{Implications of generative AI for information access}
\label{sec:cmr}
We present a reflection on the potential sociotechnical implications of generative AI, with an emphasis on LLMs, for information access.
Generative AI is still an emerging technology and our understanding of its sociotechnical impact today, and how it may evolve over time, is fairly limited.
Our treatment of this topic is therefore necessarily both incomplete and speculative.
We are informed by several recent works~\citep{bender2021dangers, weidinger2021ethical, weidinger2022taxonomy, stahl2024ethics} that attempts to map the landscape of risks and harms from LLMs.
What distinguishes our treatment of this topic relative to this previous literature is the specific focus on information access.
There has also been work on the considerations for specific applications of LLMs in IR, such as for generating direct responses to users' expressed information needs~\citep{shah2022situating}, which is relevant to our current discussion.
However, a thorough exploration of every potential application of LLMs in IR systems is beyond the scope of our current work.
Instead, we explore the implications for information access through a broader lens that encompasses considerations for content creation, content retrieval, sociopolitical power dynamics, geopolitical inequities, crowd-work, ecology, and future of IR research.
We reference relevant previous taxonomies and studies throughout this section to both support our claims and to establish meaningful connections in an attempt to present a more complete and consistent view on this topic to the reader.

We adopt the consequences-mechanisms-risks (CMR) framework proposed by \citet{gausen2023framework} to structure our presentation.
Gausen et al. introduce the CMR framework to support designers and developers of AI (and in general any computational) systems to identify and understand
\begin{enumerate*}[label=(\roman*)]
    \item the systemic consequences of developing and deploying the technology under study in the real world,
    \item the mechanisms introduced by said technology responsible for these consequences, and
    \item the corresponding risks to relevant stakeholders.
\end{enumerate*}
The framework intentionally explicates the higher-level consequences to motivate viewing the challenges through a more systemic lens.
The mechanisms, in turn, focus on more low-level system behaviors and aspects of the technology development process that contribute to the consequences and risks, and therefore represent sites for more actionable mitigation.
These consequences and mechanisms are mapped to relevant potential risks.
Through literature survey, in this work we identify the consequences, mechanisms, and risks of generative AI in the context of information access, and organize them according to the CMR framework as shown in Table~\ref{tbl:cmr}.
While we acknowledge that this list of consequences-mechanisms-risks is incomplete, we hope that it provides a summary of the sociotechnical concerns already identified in existing literature and provokes new questions for critical future research.

\begin{table}[]
\caption{Overview of potential negative consequences for information access from generative AI, the related mechanisms introduced by these AI technologies, and corresponding risks.}
\label{tbl:cmr}
\def\arraystretch{1.3}
\begin{tabular}{|p{.23\linewidth}|p{.43\linewidth}|p{.26\linewidth}|}
\hline
\textbf{Consequences} & \textbf{Mechanisms} & \textbf{Risks} \\ \hline
\multirow{6}{\linewidth}{Information ecosystem disruption (\S\ref{sec:cmr-consequence-info-ecosystem})} & Content pollution (\S\ref{sec:cmr-mechanism-pollution}) & \multirow{12}{\linewidth}{Risks to society: democracy, health and wellbeing, and global inequity (\S\ref{sec:cmr-risks-society})} \\ \cline{2-2}
& The ``Game of telephone'' effect (\S\ref{sec:cmr-mechanism-telephone}) & \\ \cline{2-2}
& Search engine manipulation (\S\ref{sec:cmr-mechanism-seo}) & \\ \cline{2-2}
& Degrading retrieval quality (\S\ref{sec:cmr-mechanism-retrieval}) & \\ \cline{2-2}
& Direct model access (\S\ref{sec:cmr-mechanism-direct-access}) & \\ \cline{2-2}
& The paradox of reuse (\S\ref{sec:cmr-mechanism-paradox}) & \\ \cline{1-2}
\multirow{3}{\linewidth}{Concentration of power (\S\ref{sec:cmr-consequence-concentration})} & Compute and data moat (\S\ref{sec:cmr-mechanism-moat}) & \\ \cline{2-2}
& AI persuasion (\S\ref{sec:cmr-mechanism-persuasion}) & \\ \cline{2-2}
& AI alignment (\S\ref{sec:cmr-mechanism-alignment}) & \\ \cline{1-2}
\multirow{3}{\linewidth}{Marginalization (\S\ref{sec:cmr-consequence-marginalization})} & Appropriation of data labor (\S\ref{sec:cmr-mechanism-labor}) & \\ \cline{2-2}
& Bias amplification (\S\ref{sec:cmr-mechanism-bias}) & \\ \cline{2-2}
& AI exploitation and doxing (\S\ref{sec:cmr-mechanism-doxing}) & \\ \cline{1-3}
\multirow{3}{\linewidth}{Innovation decay (\S\ref{sec:cmr-consequence-innovation})} & Industry capture (\S\ref{sec:cmr-mechanism-industry-capture}) & \multirow{3}{\linewidth}{Risks to IR research (\S\ref{sec:cmr-risks-research})} \\ \cline{2-2}
& Pollution of research artefacts (\S\ref{sec:cmr-mechanism-research-pollution}) & \\ \cline{1-3}
\multirow{2}{\linewidth}{Ecological impact (\S\ref{sec:cmr-consequence-ecology})} & Resource demand and waste (\S\ref{sec:cmr-mechanism-resource-waste}) & \multirow{2}{\linewidth}{Risks to environment (\S\ref{sec:cmr-risks-environment})} \\ \cline{2-2}
& Persuasive advertising (\S\ref{sec:cmr-mechanism-ads}) & \\ \cline{1-3}
\end{tabular}
\end{table}

\subsection{Consequences and mechanisms}
\label{sec:cmr-cm}
In the context of information access, we identify five potential categories of negative consequences of generative AI, and corresponding mechanisms, that we discuss next. 

\subsubsection{Consequence: Information ecosystem disruption}
\label{sec:cmr-consequence-info-ecosystem}

To reflect on the implications of generative AI on information access, we must consider the information ecosystem as a whole, and not constrain our discussion only to the application of these emerging technologies directly in IR systems.
This ecosystem includes different actors and stakeholders such as information seekers, content producers, IR systems developers, advertisers, and other sociopolitical actors.
While the information ecosystem is constantly evolving, generative AI holds the potential to significantly disrupt how each of these actors operate on their own and how they relate to other actors and stakeholders.
This potential for disruption spans across how content is produced, consumed, monetized, and used towards specific ends.
By no means do we want to imply that these plausible changes are inherently bad but the scale of potential disruptions across the ecosystem should motivate careful and thoughtful considerations before these technologies are deployed at scale.
We discuss next some the underlying mechanisms introduced by generative AI that may contribute to these disruptions.
We encourage the reader to view these mechanisms not just in isolation but to also consider how they may interact with each other and how that may impact the ecosystem over time.

\subsubsubsection{Mechanism: Content pollution}
\label{sec:cmr-mechanism-pollution}
Generative AI enables low-cost generation of derivative low-quality content at unprecedented scale.
As a consequence, synthetic AI-generated content is rapidly and very widely appearing on the web~\citep{hoel2024ai}.
On Amazon,\footnote{\url{https://www.amazon.com/}} AI-generated content includes scammy derivatives of existing publications~\citep{oremus2023he, knibbs2024scammy, limbong2024authors} and fake travel guides~\citep{kugel2023new}.
On YouTube,\footnote{\url{https://www.youtube.com/}} AI-generated video creators have targeted children~\citep{bbc2023ai, hoel2023here, knibbs2024your}.
We are also witnessing a proliferation of news websites almost entirely generated by AI~\citep{sadeghi2023rise}, which are being surfaced in search results~\citep{cox2024google} and funded by online ads~\citep{brewster2023funding}.
Even reputable publishers have reportedly published AI-generated articles under fake AI-generated author profiles~\citep{dupre2023sports}.
Beyond news, other synthetic content such as AI-generated images is starting to pollute search results~\citep{alsibai2023top, dupre2023top}.
According to another recent study~\citep{thompson2024shocking}, a ``shocking'' amount of content on the web today is machine-translated text. The promise of machine translation is that it could make more content  accessible to wider audiences. However, it also amplifies the influence of (sometimes questionable-quality) language technology choices. For example, \citet{thompson2024shocking} found that more low quality content---rather than high quality content---was machine translated into lower resource languages, likely with the goal of generating ad revenue.
Concerns have also been raised about LLMs potentially serving as ``Misinformation Superspreaders''~\citep{brewster2023next, pan2023risk} as they make it trivially easy to inundate the web with ``firehoses of falsehoods''.\footnote{\url{https://en.wikipedia.org/wiki/Firehose_of_falsehood}}
\citet{hoel2023here} points out that AI pollution of our information ecosystems is a ``tragedy of the commons''~\citep{hardin2018tragedy}.

Pollution of our information ecosystem at such scale has critical implications for people and society.
When authoring a document requires significant time and effort then the quality, style, and comprehensiveness are factors that readers may consider in deciding whether and how much to trust its content.
However, when the cost of writing an extensive article approaches zero, it becomes significantly harder for the reader to make that decision. They may not be able to distinguish between an article created based on extensive research, fact checking and thoughtful writing practices versus one generated instantly based on a short user prompt.
Furthermore, the increasing adoption of these same AI authoring tools by reputable publishers and content producers may homogenize the language and style of content on the web, making it even more difficult for readers to distinguish them from low-quality AI-generated content whose sole intent is to attract ad revenue or to mislead.
Such web pollution is also a concern for future AI models that require large web-scale datasets to train on.
Including AI-generated content in the training data for new AI models may have significant negative impact on model performance, what has been referred to as ``Model collapse''~\citep{shumailov2023curse, martinez2023towards}, ``Model Autophagy Disorder''~\citep{alemohammad2023self}, and ``Habsburg AI''.\footnote{\url{https://twitter.com/jathansadowski/status/1625245803211272194}}

\subsubsubsection{Mechanism: The ``Game of telephone'' effect}
\label{sec:cmr-mechanism-telephone}
LLMs have recently been employed in conversational search interfaces.
In systems such as Bing Copilot, the LLM has access to relevant web search results from which it can draw information to produce appropriate responses for the information needs expressed by a user.
In this scenario, the LLM performs a complex summarization task extracting relevant information from the retrieved documents to answer the search query.
In doing so, the LLM now inserts itself between the user and the retrieved web results.
This shifts the responsibility of inspecting the information in the documents and assessing their relevance, trustworthiness, and surrounding context from the user to the LLM.
Further, factual errors and inconsistencies may arise between what the LLM produces and what is in the retrieved documents.
Seeing the model through an anthropomorphic lens, these errors are sometimes referred to as ``hallucinations''.
A more technical view may see this as a noisy translation akin to the children's game of telephone.\footnote{\url{https://en.wikipedia.org/wiki/Game_of_telephone}}
Such errors, often subtle and hard to spot, may contribute to misinformation and reduce robustness of the information access system.
While the LLM-generated responses may cite relevant documents, it is unlikely that users diligently click the provided links and verify the information in the response is indeed supported by said sources.
Even if the LLM reproduces exact pieces of text from the source documents without error, taking these out of the context of the document may lead to unexpected negative consequences.
Such examples have previously been reported~\citep{varghese2021how} in context of extracted answers that search engines display on the search result pages (SERPs) as response to the user query.
These issues may become more prevalent if conversational search interfaces become a popular way to access online information.

In a more radical proposal, \citet{metzler2021rethinking} have suggested that LLMs could directly replace retrieval systems and respond directly to the user based on information in their training data.
LLMs are trained to produce statistically plausible text sequences and any semblance to an information retrieval system is likely an important mis-categorization of these models that we should be wary of~\citep{shah2022situating}.
The game of telephone effect is likely to be more intense when LLMs are expected to produce information from their training data and not just the in-context information in its input.

The interjection of the LLM between the user and the search results may have other long term effects.
These interfaces may disincentivize users from the practice of verifying information sources and make them less skilled over time at discerning online misinformation.
If users get accustomed to information being presented neatly summarized and disconnected from original sources, the critical cognitive skills necessary to distinguish between trustworthy and untrustworthy information may atrophy.

\subsubsubsection{Mechanism: Search engine manipulation}
\label{sec:cmr-mechanism-seo}
New applications of LLMs to the IR stack have exposed new attack vectors.
Prompt injection attacks~\citep{greshake2023not, liu2023prompt, liu2024automatic} that try to blur the line between instructions and data have garnered specific interest.
In these types of attacks, website owners may inject what looks like instructions to the LLM.
When such documents are retrieved and included in the input of the LLM as augmentation, the LLM may mistake the injected prompt in the document content and be vulnerable to manipulation.

Recently, LLMs have also found application in relevance labeling for search~\citep{thomas2023large}.
It is not well understood yet whether this may make the search engine vulnerable to improper ranking manipulation by website owners and search engine optimization experts.
For example, one may employ the same, or similar, LLMs to reproduce the labeling scheme externally and then adapt their website content and design to achieve undue high predicted relevance against queries to rank higher on SERPs.

Other attack vectors may include using LLMs to create effective content farms at low cost to manipulate the ranking of web results, or even use LLMs to artificially simulate users interacting with the search system to fake clicks and other user behavior signals, such as reformulations, that search engines depend on.

\subsubsubsection{Mechanism: Degrading retrieval quality}
\label{sec:cmr-mechanism-retrieval}
LLM usage can negatively impact search result quality in a number of (indirect) ways.
LLMs can contribute to new attack vectors, but more worryingly, in some cases the negative effect may be a result of the LLM behaving exactly as it is supposed to.
For example, one potential consequence of using conversational search interfaces,
is that the quality of feedback from user behavior signals on SERPs may significantly degrade.  Historically, users of commercial web search engines have given search systems noisy implicit feedback through clicks and other actions on SERPs. These actions are one of the key secret sauce of any modern search systems.

However, conversational interfaces may discourage direct user clicks on web results and at best provide much weaker satisfaction signal that may be gleaned from the users' next utterance in the conversation. 
This over time may negatively impact the underlying retrieval quality.
This makes it important to invest in methods that can infer user satisfaction with high certainty from the natural language conversations.
However, methods for such signal interpretation are not yet at the level necessary to mitigate these impacts.

In conversational search interfaces and other applications, such as Microsoft Copilot for M365~\citep{mehdi2024bringing, warren2024microsoft}, the LLM may conduct the search on the user's behalf.
In this process, the LLM generates search queries.
If these queries differ from those that are likely to be submitted by users then the underlying search system needs to optimize itself for both real user queries and LLM-generated queries.
This may have consequences that are not yet well understood.
Optimizing the search system directly to improve the LLMs natural language responses may also have unforeseen outcomes, especially in light of the fact that what makes for a good result set for retrieval-augmentation is not yet fully understood~\citep{cuconasu2024power}.

\subsubsubsection{Mechanism: Direct model access}
\label{sec:cmr-mechanism-direct-access}
Another important consideration is the implications of open foundation models~\citep{kapoor2024societal}.
While centralized systems have their own negative implications, as discussed in \S\ref{sec:cmr-consequence-concentration}, open access generative AI models without any access moderation also pose certain challenges.
For example, there are many classes of harmful intents that systems should refuse to respond to.
This may include search queries seeking information on methods to self-harm or cause harms to others, or requests to generate harmful (and sometimes illegal) content such as child sex abuse material (CSAM) or non-consensual intimate information (NCII).
Publicly accessible LLMs trained on large web corpora may produce such irresponsible content in the absence of moderation.
Even if a model is trained to not respond to certain classes of queries, it is likely that there will be leakage, and the safety alignment may also be compromised if the model is further finetuned~\citep{qi2023fine}.
Such leakage may also happen in the context of traditional search systems.
However, in the latter case, all queries are typically logged, allowing for post-hoc analysis and identification of critical gaps in the moderation system.
Unfortunately, no such mitigation is possible once these generative AI models are released into the wild.

\subsubsubsection{Mechanism: The paradox of reuse}
\label{sec:cmr-mechanism-paradox}
Content producers and information access technologies are critically inter-dependant~\citep{mcmahon2017substantial, vincent2018examining}.
Websites such as Wikipedia,\footnote{\url{https://www.wikipedia.org/}} StackExchange,\footnote{\url{https://stackexchange.com/}} and Reddit\footnote{\url{https://www.reddit.com/}} produce critical content that is surfaced by information access platforms (\eg, web search engines) and contribute to making these platforms significantly more useful to their users.
In return, these platforms have historically sent traffic back to the websites that contributes to their increased readership, subscriptions, and monetization.
However, when search platforms stop directing traffic back to websites---\eg, by instead surfacing relevant content directly on the search result pages (SERPs)---the relationship becomes less symbiotic towards the content producers, a phenomenon \citet{taraborelli2015sum} termed the ``paradox of reuse''.

The application of LLMs as conversational information access interfaces is likely to significantly intensify this problem.
For example, LLMs such as ChatGPT\footnote{\url{https://chat.openai.com/}} and Google Gemini\footnote{\url{https://gemini.google.com/app}} may gobble up large quantities of content from websites as part of their training data and later regurgitate the same information without any attribution back to the sources.
Even when models summarize information from multiple online sources with attribution---\eg, Bing Copliot\footnote{\url{https://www.bing.com/chat}}, they typically de-emphasize the references and reduce the likelihood of the searcher clicking through to the source websites as compared to the classic ten-blue-links interface.
There is evidence~\citep{del2023large} to suggest that this phenomenon is already happening at scale and is jeopardizing the ``grand bargain at the heart of the web''~\citep{hays2024ai}.

\subsubsection{Consequence: Concentration of power}
\label{sec:cmr-consequence-concentration}

\begin{quote}
  ``We may have democracy, or we may have wealth concentrated in the hands of a few, but we can't have both.''
  \begin{flushright}
  -- Louis Brandeis
  
  \emph{As quoted by~\citet{lonergan1941mr}}
  \end{flushright}
\end{quote}

Technology shapes and is shaped by the sociopolitical power structures within which it exists.
The 2024 edition of the World Economic Forum’s Global Risks Report~\citep{wef2024world} lists ``technological power concentration'' as one of the top global risks for the coming decade and as the biggest upward mover in their annual ranking of global risks compared to the previous year. Deliberation on the social consequences of any technology must therefore include critical consideration of how the technology, and general narratives about said technology, shifts power and re-architects and codifies structures of hierarchy and control.
In this context, the politics and values of those in power to oversee what and how technology is built or regulated, especially when they reinforce hierarchy and authoritarianism (\eg~\citep{gebru2023eugenics, lafrance2024rise, duran2024}), becomes important to consider.

A report~\citep{kak2023ai} from the research institute AI Now\footnote{https://ainowinstitute.org/} similarly asserts ``the concentration of economic and political power in the hands of the tech industry---Big Tech in particular'' as the core challenge posed by AI.
They further note that not just the technologies but the narratives (both the hype and the fear-mongering) around them questionably bolster claims of ``foundational'' advancements and their unassailable equivalence with scientific progress.
These concerns are complemented by the discourses within the AI community, such as the observations by \citet{birhane2022values} that the prominent values expressed and operationalized in top cited AI papers generally have implications in support of centralization of power.
Even if platform owners act accountably to civil society, the concentration of power and control in their hands makes them vulnerable to other actors, such as autocratic governments, and allows that power to be potentially abused for oppressive and harmful intents. 

The popularization of generative AI can concentrate that power within large companies, since they emerge as some of the only institutions with the resources to develop and deploy these technologies~\citep{khanal2024and}.
The application of these technologies for information access may contribute to further concentration and growing inequities of wealth and power; we discuss three mechanisms in the context of generative AI that may contribute to concentration of power and control.

\subsubsubsection{Mechanism: Compute and data moat}
\label{sec:cmr-mechanism-moat}
The development of generative AI is heavily reliant on the availability of large swaths of training data and large-scale computing power for training and deployment.
Only a handful of institutions, largely in the private sector, own and control these necessary resources while simultaneously evangelizing AI as crucial geopolitical leverage and critical social infrastructure~\citep{kak2023ai}.
Increased access to these models has sometimes been touted as potential paths to mitigation~\citep{euaiactoss, solaiman2023gradient}, where access may range from being heavily restricted over API to ``open weight'' models~\citep{liesenfeld2024rethinking}.
The ability to download  models with their learned parameters allows others to further adapt for their own applications and opens the door to more meaningful analysis and audit of these models.
However, such ``open access'' also leads to severe limitations that we should recognize.
The availability of the trained models does little to challenge the predominant visions put forth by large technology companies of what AI fundamentally should look like.

One potential direction would be to dismantle the data and compute moat by turning them over from private ownership into public infrastructure for independent researchers and developers and those affiliated with smaller institutions.
This also illustrates the importance of existing institutions such as archives, libraries and universities that have reliable, historical data.
The availability of public computer infrastructure would allow a broader set of developers to participate in the reimagination and development of diverse approaches to AI and not merely being forced to be satisfied with critiquing and finetuning artefacts produced by other institutions. However, there is no guarantee that without careful planning and incentives, a proliferation of smaller projects will lead to transformative new or more sustainable results.

Democratizing the control over computational resources provides a mechanism of checks and balances on the future directions of AI systems, and may allow for challenges to popular narratives and expectations about generative AI such as exponential growth in model size over time. Infrastructure is however also bound to the particular governing system, and local underlying goals and processes. Larger investments in existing research institutes, or new alternative companies or non-profits might in certain cases lead to faster results. 

 Similarly, the research community would benefit from easier access to industry models and APIs for critical studies and auditing. However, access to models or APIs alone is significantly limiting unless that access is also extended to the user-facing systems in which these technologies are deployed. The corresponding instrumentation data would provide context on how these systems are used by people and potential consequences. This can lead to practical privacy and security questions for platform teams. Practical support for decision making and for example the creation of standards to de-risk those concerns can help alleviate some of those concerns.

\subsubsubsection{Mechanism: AI persuasion}
\label{sec:cmr-mechanism-persuasion}
There is an emerging recognition of the dangers of \emph{AI persuasion}~\citep{burtell2023artificial, carroll2023characterizing, park2023ai, el2024mechanism}, which \citet{burtell2023artificial} define as ``a process by which AI systems alter the beliefs of their users''.
AI systems may persuade users by appealing to their reason and argument, or by using their cognitive biases and heuristics~\citep{el2024mechanism}.
\citet{el2024mechanism} identify six mechanisms of generative AI persuasion---namely
\begin{enumerate*}[label=(\roman*)]
    \item trust and rapport,
    \item anthropomorphism,
    \item personalization,
    \item deception and lack of transparency,
    \item manipulative strategies, and
    \item alteration of choice environment
\end{enumerate*}---and corresponding model features that contribute to these mechanisms.
In the context of information access and advertising, these capabilities of generative AI can be powerful tools to hyper-target users and steer their behaviors.

Modern online information access and communication platforms monetized with targeted advertising have been said to usher in an age of surveillance capitalism~\citep{zuboff2023age, zuboff2019surveillance}.
Information access systems increasingly collect detailed user behavior data that allow them to build accurate user profiles for audience targeting.
There is strong evidence that people are more likely to consume information that opposes their own personal views and beliefs when the it employs language similar to their own political leanings~\citep{yom2014promoting}.
So, combining users' private preferences and behavioral data with the capabilities of generative AI to produce persuasive language could create worrying tools for mass behavioral manipulation.
The impact of such pervasive \emph{algorithmic nudging}~\citep{mohlmann2021algorithmic} may be further pronounced over longer time periods from continuous interactions between the user and the system.
Putting these capabilities in the hands of online platform owners, which typically tend to be large multinational for-profit institutions with largely hierarchical non-democratic internal governance structures, poses serious risks to functioning of democratic societies. At the same time, platforms must make decisions about what is acceptable on their platforms to avoid negative user experiences, spam, unwelcoming behavior, and other negative occurrences beyond those outlined in legal compliance alone. Platforms moderate content posted or accessible through the platform~\citep{gillespie2018custodians} and in doing so they unavoidably impose implementations of values on their users, or the values incentivized by, say, advertising needs or other business model related motivations. 
For ads, this may mean an incentive to use generative AI to produce hyper-targeted highly-personalized persuasive advertisements which convince users to make certain buying decisions.
For content, when platforms optimize for increased user engagement, they may knowingly or unknowingly incentivize generative AI models to be producing highly charged content, such as ``rage-bait''~\citep{hom2015rage}, because it tends to be more persuasive and engaging.

\subsubsubsection{Mechanism: AI alignment}
\label{sec:cmr-mechanism-alignment}
To prevent generative AI models from producing harmful and offensive content, recent research has focused on how to align model outputs with ``human values''~\citep{kasirzadeh2023conversation, tamkin2021understanding, russell2015research, gabriel2020artificial, gabriel2021challenge}.
Approaches such as reinforcement learning from
human feedback (RLHF)~\citep{christiano2017deep, ziegler2019fine} have been effective in limiting certain types of problematic content from being produced.
However, this approach presupposes some notions of desirable values and puts the burden of determining and enforcing them on the shoulders of platform / model developers.
Any notions of universal values that might determine what type of content these models should generate---or, not generate~\citep{urman2023silence}---is highly contested~\citep{prabhakaran2022human, birhane2019algorithmic, jobin2019global, png2022tensions, sambasivan2021re}.
Placing these decisions in the exclusive domain of the platform developers, especially in the absence of democratic and civil society oversight, further concentrates power and responsibility. This is not an argument against content moderation itself but against the centralization of control over it without civil oversight or broader societal participation. As a pragmatic example, platforms may not necessarily have the necessary knowledge in-house, making it imperative for them to make successful connections to outside expertise.

\subsubsection{Consequence: Marginalization}
\label{sec:cmr-consequence-marginalization}
Generative AI, both in its process of development and in its deployment in the context of information access, can marginalize groups and individuals by diminishing their value, power, and well-being.
Next, we discuss some the mechanisms that may contribute to this.

\subsubsubsection{Mechanism: Appropriation of data labor}
\label{sec:cmr-mechanism-labor}
\citet{li2023dimensions} define \emph{data labor} as ``activities that produce digital records useful for capital generation''.
The term encompasses both witting labor activities---as in the case of crowdwork~\citep{altenried2020platform}, peer production~\citep{tarkowski2023how, tarkowski2023stewarding}, and content moderation~\citep{gillespie2018custodians}---and unwitting activities such as user behavior data and other data generated when users interact with and participate on the platforms.
Data labor also encompasses the creation of artefacts by writers~\citep{cohan2023ai, coldewey2023thousands}, artists~\citep{vincent2022shutterstock, vincent2023ai}, and programmers~\citep{vincent2022lawsuit} etc. outside of the AI development process that are nonetheless extracted from the web and fed in as training data to generative AI models.
Appropriation of data labor in this context includes both
\begin{enumerate*}[label=(\roman*)]
    \item the uncompensated appropriation of works by writers, authors, programmers, and peer production communities like Wikipedia~\citep{cohan2023ai, coldewey2023thousands, vincent2023ai, vincent2022lawsuit, shrivastava2023openai, burke2023biggest, burke2024generative, vincent2021github, vincent2020dont, appel2023generative, marr2023generative, chesterman2024good, chayka2023ai, gertner2023wikipedia, vincent2023chatgpt}, and
    \item under-compensated crowdwork for data labeling that has been instrumental in the development of these technologies~\citep{perrigo2023exclusive, williams2022exploited, tan2023behind, altenried2020platform, hao2022ai, xiang2023openai, hao2023cleaning}.
\end{enumerate*}

It is particularly harmful when technology developed on appropriated labor is then employed to displace and automate the jobs of those whose labor was appropriated~\citep{anguiano2023hollywood, coyle2023hollywood, vincent2021github}.Introduction of such automation may involve vicious cycles of perceived skill-transfer from people to AI models whereby professional jobs are replaced by corresponding lesser-paid gigified equivalent as auditing and editing of model outputs only~\citep{gordon2023co}. Proprietary AI model capabilities may then continue to improve by learning from workers' inputs, while workers progressively lose their economic value and power, or are even relegated into the role of \emph{moral crumple zones}~\citep{elish2019moral}.

This is a critical challenge in the context of information access because
\begin{enumerate*}[label=(\roman*)]
    \item the devaluation of writers and artists have direct implications for the quality of content on the web, and
    \item these automated content generation tools are starting to get incorporated directly in information access platforms~\citep{pierce2023you}.
\end{enumerate*}
Similar concerns of commodification and appropriation have also been raised in other information and knowledge access contexts such as in the enterprise~\citep{gausen2023framework}.

\noindent\emph{AI for me, data labor for thee.}
Another pernicious aspect of AI data labor dynamics discussed in the literature is how they can mirror and reify racial capitalism and coloniality, employ global labor exploitation and extractive practices, and reinforce the global north and south divide~\citep{hao2022artificial, birhane2020algorithmic, o2023heart, klein2023ai, couldry2019data, muldoon2023artificial, tacheva2023ai}. While worldwide jobs might be created in certain cases, the workers are typically low-paid and deprived of any share of the profit made from technologies built with their labor.
These dynamics encompass accruing the benefits of generative AI to privileged populations, while data labor is relegated to already marginalized populations, for example in the global south.
Communities that significantly contribute to AI data labor may even find their own linguistic styles being labeled AI-ese~\citep{hern2024techscape} and being forced to repeatedly prove their own humanity~\citep{dhawan2023universities, mathewson2023ai}.
Attempts to bridge the global north-south data gap also in turn may further intensify data extractive practices in the global south~\citep{coffey2021maori}.

\subsubsubsection{Mechanism: Bias amplification}
\label{sec:cmr-mechanism-bias}
LLMs and other generative models reproduce and amplify harmful biases and stereotypes from their training datasets~\citep{bolukbasi2016man, caliskan2017semantics, gonen2019lipstick, blodgett2020language, bender2021dangers, abid2021persistent} which can lead to allocative and representational harms~\citep{crawford2017trouble}.
Harms may also materialize from \emph{demographic blindness}~\citep{gausen2023framework} when the model (or the system it is embedded in) treats different individuals and groups as alike when, in fact, it is unwarranted.
Examples may include the handling of certain languages as one homogeneous entity without regards for sociolects or dialects~\citep{blodgett2016demographic} or holding different perspectives as equally valid without considerations for historical context or structural dynamics of power.
These biases are concerning in the context of information access systems that are responsible for supporting informed citizenry and functioning democracies, health literacy, and knowledge production among other societal needs.

\subsubsubsection{Mechanism: AI exploitation and doxing}
\label{sec:cmr-mechanism-doxing}
\emph{``AI doxing''} can describe the act of leaking people's private information by an AI system.
\citet{weidinger2021ethical} note that this may be caused by models leaking private information (\eg, address and telephone number) present in their training data~\citep{carlini2021extracting} or when these models are employed to predict people's sensitive attributes (\eg, political and sexual identities) based on what is known about them publicly~\citep{kosinski2013private, park2015automatic, quercia2011our, youyou2015computer}.
Private information in the training data is a challenge even if the datasets have been sourced from the public web because models may continue to regurgitate that information after it has been removed from the web, or bypass safety measures that would prevent such information from surfacing through web search---\eg, the information may be protected by robots.txt that blocks popular search crawlers but misses crawler bots that specifically collect data for AI model training.
In many contexts, applications of these models to predict people's private information may be based on shaky scientific grounds~\citep{aguera2017physiognomy, vincent2017invention}, to put it mildly.
However, such applications may still contribute to serious harms and discrimination regardless of their accuracy as long as some people are convinced of their predictive power and employ them to marginalize others.
AI doxing may also take other forms such as reverse-image-search~\citep{baio2024most}, a functionality supported by some search engines, that may be abused for stalking and harassment.
In turn, exploitative materials produced with GenAI (such as deep fake revenge porn, or CSAM) might be amplified.

\subsubsection{Consequence: Innovation decay}
\label{sec:cmr-consequence-innovation}
Generative AI may find innovative new applications in information access.
However, the excitement around these technologies and the significant investments from industry, government, and academia on corresponding research and development have broader implications for IR research.
Next, we discuss some of the mechanisms associated with the research and development of generative AI that may potentially throttle innovation in information access technologies.

\subsubsubsection{Mechanism: Industry capture}
\label{sec:cmr-mechanism-industry-capture}
The compute and data moat that concentrates power in the hands of big tech, as discussed earlier in \S\ref{sec:cmr-mechanism-moat}, also creates significant barriers to entry for academic research.
These barriers limits academic AI research to a handful of institutions that have the necessary means and connections to industry who provide access to compute and data resources to incentivize research in areas of their economic interests.
Academics who want to contribute to research on large scale AI systems or critique their sociotechnical impacts are pressured to play well with institutions holding monopolistic control over compute, data, and systems~\citep{murgia2019ai}.
Access to ``open access'' models---without the compute and data necessary to build them from scratch---allows academic researchers to invest in finding more effective applications of these technologies that serve industry interests, but not to reimagine / rearchitect them to in radically different ways.
Students and other academics who may someday want to work in industry are shepherded into integrating themselves into this homogenized research agenda.

Such ``industry capture''~\citep{whittaker2021steep} allows for inordinate influence of the sociotechnical imaginaries\footnote{\citet{jasanoff2015dreamscapes} define \emph{sociotechnical imaginaries} as ``collectively held, institutionally stabilized, and publicly performed visions of desirable futures, animated by shared understandings of forms of social life and social order attainable through, and supportive of, advances in science and technology''.} of profit-driven
corporations over for example academic researchers~\citep{mitra2024search}. This can thwart research that may not be immediately monetizable or challenges the status quo of power concentration, and complements the ``regulatory capture'' by bigger tech companies~\citep{liu2023letter, schaake2021big, asokan2024uk}.
As \citet{mitra2024search} asks: ``\emph{Whose sociotechnical imaginaries are granted normative status and what myriad of radically alternative futures are we overlooking?}''
Narratives of the inevitability of these technologies that are hyped up to be both transformative forces for society and simultaneously posing existential risks for humanity (often purported by the same actors) only bolster their imagined importance to accumulate increasing global investments, including from governments.
Researchers who care about sociotechnical impact and ecological sustainability are busy with enumerating the harms of rapidly emerging new AI technologies and chasing potential mitigations instead of having the full means to imagine and develop systems for social good.
While industry practitioners can contribute to both identifying new research challenges grounded in real-world systems and practical methods to mitigate some of the risks of emerging technologies, it is imperative that we create avenues for increasing independent research, while preserving the benefits of various modes of industry-academia collaborations. 

Even as the grounded risks from these technologies (such as those discussed here) gather consensus from academic communities and civil society, it can be difficult to create space for alternative ways of development that are perceived as ``slowing down''.
Critical research on sociotechnical harms of AI is also under risk when attempts are made to shift attention from concerns about real harms to marginalized people today to unsubstantiated imagined future concerns~\citep{gebru2023eugenics, gebru2022effective}.
Calls for regulations to address these imagined future harms~\citep{gebru2023statement} further detract from real progress and contribute to reinforcement of monopolistic powers of those who have already added these technologies to their arsenals.
This has led some sociotechnical researchers in AI to explicitly draw attention to how these systems shift power (\eg, \citep{kalluri2020don, blodgett2020language, miceli2022studying, gausen2023framework}), and to prioritize research guided by alternative visions for sociotechnical futures grounded in universal emancipation and social justice~\citep{mitra2024search}.
It is thus important that access to investments to enable development is also available to those trying to not only mitigate existing systems' harms, but also develop new avenues, including work on social good and new business models.


As generative AI starts to accumulate the lion's share of research investments, it may starve out other areas of information access research.
Generative AI has had exciting but limited deployments in information access systems today.
There are significant open challenges to making these models broadly useful, including but not limited to concerns of potential sociotechnical harms.
There is a risk that if these challenges are not mitigated in spite of the extensive resources already invested on them at present, there may be calls for even larger investments in future prompted by the sunk cost fallacy.\footnote{\url{https://en.wikipedia.org/wiki/Sunk_cost\#Fallacy_effect}}
It would be astute for the IR community to consciously continue to invest in research on systems and applications that societies need beyond what existing AI technologies make plausible~\citep{shah2022situating, mitra2024search}.

\subsubsubsection{Mechanism: Pollution of research artefacts}
\label{sec:cmr-mechanism-research-pollution}
Risks to academic research from generative AI may also emerge through the applications of generative AI models in IR scholarship---\eg, for authoring scientific papers and peer reviewing.
There is evidence that researchers in computational sciences are already leveraging these tools~\citep{liang2024monitoring}, sometimes with hilariously terrible outcomes~\citep{pearson2024scientific}.
While the use of language models for light editing may (eventually) fall within the norms of socially acceptable behavior in research, their application in scholarship does raise concerns of plagiarism and scientific inaccuracies.
This is an area that currently has more questions than answers and the IR community would benefit from proactively considering potential implications of this trend on future IR research. 

\subsubsection{Consequence: Ecological impact}
\label{sec:cmr-consequence-ecology}
Another important consequence of generative AI is its impact on the environment.
In this context it is important for us to consider the direct environmental cost of developing and deploying generative AI systems at scale as well as the potential impact of these technologies on the climate change discourse online.

\subsubsubsection{Mechanism: Resource demand and waste}
\label{sec:cmr-mechanism-resource-waste}
The ecological cost of deep learning models has been a subject of much concern and debate in the AI community~\citep{strubell2019energy, bender2021dangers, patterson2021carbon, bommasani2021opportunities, wu2022sustainable, dodge2022measuring, patterson2022carbon, kanungo2023green, berreby2024use}.
Similar concerns have also been raised within the IR community with respect to the application of these models for information access~\citep{scells2022reduce, zuccon2023beyond}.
By some estimates, the computing power being utilized for deep learning research has been doubling every 3.4 months since 2012~\citep{cantrelltrue}.
In the US, data centers consumed more than $4\%$ of the total national electricity in 2022, and that number is projected to grow to $6\%$ by 2026~\citep{halper2024amid}.
Another study~\citep{belkhir2018assessing} estimates that by 2040 Information and Communications Technology industry on the whole will account for $14\%$ of global emissions.
Beyond emissions, data centers' water consumption is also raising alarm bells~\citep{li2023making, ren2023much, guerrini2023ai, hao2024ai, gordon2024ai, gupta2024ai, criddle2024ai, naughton2024ai}.
By 2027, global AI demand may be responsible for withdrawal of 1.1--1.7 trillion gallons of fresh water annually~\citep{li2023making, hao2024ai}.
Serious concerns also revolve around the rising levels of electronic waste~\citep{khattak2023environmental}.
Even as we make progress in reducing the ecological cost of training and deploying the current AI models, we risk encouraging the development of even larger models and their wider deployment worsening the overall ecological impact (\ie, Jevons paradox).\footnote{\url{https://en.wikipedia.org/wiki/Jevons_paradox}}

\subsubsubsection{Mechanism: Persuasive advertising}
\label{sec:cmr-mechanism-ads}
Generative AI may not only negatively impact the environment through increasing demand for natural resources and increasing generation of waste, but may also supercharge climate change disinformation~\citep{galaz2023ai, foe2024report, sustainablebrands2024report, disinformation2024artificial, Speare2024generative, corbett2024report}.
For example, the fossil-fuel industry may attempt to sway public opinion through advertising that leverages generative AI's persuasion capabilities discussed in \S\ref{sec:cmr-mechanism-persuasion}.
Persuasive advertising may also be employed by other environmentally-unfriendly business models like fast-fashion~\citep{coleman2023ai}.
While the direct ecological cost of generative AI justifiably garners lots of attention, its potential impact on related online discourse also deserves scrutiny.

\subsection{Risks}
\label{sec:cmr-risks}
We categorize the risks of generative AI broadly to our society, to IR research, and to the environment.
We map the first three consequences discussed earlier in this section---\ie,
\begin{enumerate*}[label=(\roman*)]
    \item Information ecosystem disruption (\S\ref{sec:cmr-consequence-info-ecosystem}),
    \item concentration of power (\S\ref{sec:cmr-consequence-concentration}), and
    \item marginalization (\S\ref{sec:cmr-consequence-marginalization})---and
\end{enumerate*}
their corresponding mechanisms as potentially contributing to the risks to society.
We further map the last two consequences---\ie,
\begin{enumerate*}[label=(\roman*)]
    \setcounter{enumi}{3}
    \item Innovation decay (\S\ref{sec:cmr-consequence-innovation}) and
    \item Ecological impact (\S\ref{sec:cmr-consequence-ecology})---to
\end{enumerate*}
the risks to IR research and the environment, respectively.

\subsubsection{Risks to society}
\label{sec:cmr-risks-society}
Information access is a critical need of any democratic society and a necessary ingredient for social transformation~\citep{higgins2013information, polizzi2020information, goldstein2020informed, correia2002information, gonzalez2021better}.
It is also a social determinant of economic progress~\citep{yu2002bridging, mutula2008digital} and health~\citep{moretti2012access}.
Disruptions to the information ecosystem bears potentially grave risks to most aspects of our social lives.
A confluence of the pandemic~\citep{scientist2021covid, taylor2022covid, centers2022cdc}, rising global conflicts~\citep{taylor2023historic, un2023highest}, and escalating climate catastrophes~\citep{parmesan2022climate, ipcc2013physical, poynting2024confirmed} are pushing the world towards precarious instability.
Our information ecosystems are already struggling under the weight of misinformation and disinformation that in this critical moment is eroding public trust in online platforms, institutions, and each other.
It is imperative that researchers and developers of information access systems prioritize safeguarding social interests and be vigilant in considering potential risks of disruption and ecosystem collapse when integrating generative AI technologies in the IR stack.
This includes identifying the necessary conditions under which these technologies can be safely deployed and developing practical safeguards and alternatives. 

The risks to society are not just from potential disruptions of the information ecosystem, but also from how these technologies simultaneously concentrate power away from those at the margins of society.
As institutions that develop and operate these technologies are themselves beneficiaries of this concentration, we need democratic oversights.
If technologies further exacerbate already worsening wealth and power inequities, this additionally may pose severe threats to democratic institutions and human rights. There is an opportunity cost of not re-imagining information access in light of sociotechnical ambitions of human emancipation, culture, and knowledge production, instead of being constrained solely by what these emerging technologies make plausible and the homogenized visions put forth by institutions who wield these technologies~\citep{mitra2024search}.

\subsubsection{Risks to IR research}
\label{sec:cmr-risks-research}
IR research can suffer from a confluence of different factors including the distancing of academic researchers from the data and compute they need to do their work and how narratives about the inevitability of AI technologies shapes what computational research gets funded.
The concentration of access to the networks around these technologies in a subset of institutions shapes what is considered ``foundational'' or even ``AI''.
Research on generative AI should not be performed only in the context of corporate economic interests while academia is hollowed out and prevented from exploring radical new methods that challenge the status quo.
This risk of homogenization of academic research agendas and the opportunity cost of not exploring more diverse approaches to online information access can have material consequences. Instead, the IR community must be empowered with both the space and the resources necessary to explore a diversity of these visions and critique dominant narratives. IR research should have a plurality of work, which includes work with access to industry to change current practices. However, we especially also need to ensure that not all IR research is simply an extension of industrial system development and risk the demise of fundamental research on alternative avenues. 

\subsubsection{Risks to environment}
\label{sec:cmr-risks-environment}
Information access provides one of the large scale application settings for generative AI.
However, the impact of such wide-scale deployment of these technologies on the impending climate crisis should be a critical consideration. Climate costs pose substantial existential risks for ecosystems and people, in more direct ways than some other ``existential risks'' that lack adequate scientific basis but have nonetheless been popular discourse in some parts of the AI community.  This means both choosing what to deploy, and investment in methods to mitigate negative impacts that build on existing environmental work. As we discussed in \S\ref{sec:cmr-consequence-ecology}, these concerns include not just the ecological cost of developing and deploying generative AI technologies but also their impact on online discourse on societal priorities.
\section{Methods to evaluate risks and impact}
\label{sec:evalandmitigation}

\subsection{Evaluating the impact of generative IR applications}
Evaluating the impact of generative IR applications requires methods, as do data-informed interventions to steer that impact. Creating an LLM-based demo has become exceedingly easy. Understanding the impact of a system when it gets used in real life contexts, and getting to a high quality experience for a wide variety of users, is much harder. Standards for impact assessment have not kept up a similar pace as tech developments. Klaaf points out the need to carefully consider the differences in value alignment of the goals of a system, and safety considations, harms and risks~\citep{khlaaf2023toward}. A wide range of online, offline, and human-assisted evaluations are possible -and necessary- to get a full sense of the impact of a system. 

There are a number of frameworks that can provide helpful starting points for evaluating the impact of generative IR applications, and potential quality or safety improvements. Not surprisingly however, they can measure quite different aspects of a system and its underlying models. Distinctions have to be made between evaluating a model, a system, or a technology as a whole. For example, standards for foundation model evaluations might not take into account the impact of a system that uses such a model (or a combination of models) in a specific application context. 

Measurement and interventions are possible at every stage of the development life cycle of products, and their underlying models and data. In this regard, general insights around for example harm mitigation interventions being possible throughout the Machine Learning life cycle \cite{Suresh21} also apply to generative IR. To improve quality and safety, we need to be able to operationalize and measure the impact of potential interventions. This includes evaluations on aspects of that might be both system performance issues, but are also of societal importance, e.g., harmful/toxic output, hallucination, and differing model performance across languages/demographics. 

\subsection{Threat identification, assessment, and modeling}
When the emergence of a new technology or application becomes apparent, the assessment of whether this poses risks or opportunities within specific domains poses a challenge. Before development of a system, threats and opportunities can be identified. 
As \citet{kapoor2024societal} point out, it’s crucial not to evaluate the risks and impact of new systems in isolation, but rather in comparison with existing technologies. For example, the impact of usage of foundation models in \textit{search} should be compared to existing web \textit{search}. For this purpose, Kapoor et al., present an evaluation framework focus on marginal risks, applied to Open Foundation Models. Their framework is based on threat identification work from cybersecurity and consists of six steps necessary to demonstrate such marginal risk. These steps are: 1) threat identification, 2) evaluating existing risk absent open foundation models, 3) considering existing defenses absent open foundation models, 4) evidence of marginal risk of open foundation models, 5) ease of defending against new risks, and 6) outlining uncertainty and assumptions. Note that this framework does not set exact assessment criteria, but rather defines the steps to get to such evaluations.

In practical settings, this might mean having to select standards for the development process (e.g. emerging standards from organizations such as NIST~\citep{nist2024ai} or ISO~\citep{iso2024ai}, company-specific standards such as Microsoft's Responsible AI Standard v2 General Requirements~\citep{microsoft2022responsible}, or following new (local) legal requirements). However, mapping out potential consequences and identifying mechanisms that introduce risks in the specific context of a system needs to go much further. How to disrupt potential negative mechanisms in order to mitigate those risks requires gauging a wide range of consumer-side impacts \cite{Ekstrand2024}, but also wider societal impacts. That includes frameworks focused on worker consequences \cite{gausen2023framework}, or practical methods focused on reducing the (legal) risks of using certain types of copyrighted or restricted training data vs. expected performance gains~\citep{min2023silo}.

\subsection{Evaluation during model development}
\subsubsection{Model benchmarks vs. actual system context}
LLM benchmarks are widely used to compare the quality and safety progress made by new \textit{model} releases, resulting in model leaderboards on different scenarios. The Stanford HELM~\cite{StanfordHELM} leaderboard for example shows the performance of different LLM models on benchmarks, and these benchmarks include societal impact and bias-related measures. Their HELM (‘holistic framework for evaluating foundation models’) framework \cite{liang2023holistic} uses scenarios, and measures seven metrics. Those are accuracy, calibration, robustness, efficiency, but also more social impact-oriented fairness, bias, and toxicity. Each scenario focuses on one use case, and consists of a dataset of instances, such as the LegalBench set of legal reasoning tasks \cite{guha2023legalbench}, or medical board exam problem sets \cite{JinMedical2021}.
The larger BIG-bench (“Beyond the Imitation Game benchmark”) \cite{srivastava2023beyond} consists of 200+ tasks, contributed by hundreds of authors at a variety of institutes. More specific benchmarks for trustworthiness such as DecodingTrust, in turn focus on subsets such as toxicity, stereotyping, adversarial and out-of distribution robustness, privacy, machine ethics, and fairness~\cite{wang2024decodingtrust}, while for example the much more specific recurring TREC Fair Ranking track competitively evaluates systems according to how fairly they  rank documents on a specific test task \cite{trec-fair-ranking-2021}.

Paradoxically, while these benchmarks include aspects of societal impacts such as bias and toxicity, they do not necessarily cover the aspects that matter most in a specific application context in practice. Benchmarks are generally geared towards structured comparisons between models, \textit{not} towards evaluating end-user applications in practice. This means that they may not be particularly suitable for a specific application and the people involved in its usage. In addition, using such large benchmarks can be quite resource intensive, making ‘lite’ versions necessary that are less comprehensive. Both Helm and BIG-Bench are also implemented as Lite versions. However, the evaluation differences that arise from specific, lighter implementations of benchmarks can significantly impact model comparison results \cite{srivastava2023imitation}. This makes it necessary to go beyond these benchmarks, and ensure suitable evaluations for the application at hand to avoid deriving conclusions about safety or responsibility devoid from actual application concerns. 

\subsubsection{Combining IR and generative AI evaluation metrics}
It is challenging that standards for measuring societal impact, including  bias, fairness and etc. are yet scarce in IR \textit{product} settings. For example, \citet{Smith2023} provide an overview of different metrics available for evaluating bias and fairness in recommendation systems, and the challenges practitioners face when choosing between them. In some cases, it may be more appropriate to for example focus on ‘traditional’ performance and accuracy metrics, but study the performance and subsequent quality of experiences for different groups of people by segmenting/slicing results by group. This approach assumes the ability to define relevant groups, or relies on more advanced methods to find clusters that may---or may not---have significant differences in performance or quality.

Specific methods might also be necessary to match new techniques. For example, Retrieval Augmented Generation (RAG) might be used to include more reliable information in a specific domain and reduce hallucinations in an LLM setting. However, RAG does not necessarily fully solve every hallucination-related issue. Specific frameworks that fit an application context are still necessary to evaluate these techniques and their actual impact on aspects such as factuality within that context. One example is \citet{saadfalcon2024ares}, who present an evaluation framework, ARES, for RAG-assisted Question \& Answering settings. This framework uses three evaluation scores: context relevance of the retrieved information, answer faithfulness (the answer's grounding in the retrieved context), and answer relevance to the question asked.  These are similar to IR-evaluations, but might need adjustment to the setting at hand, and datasets used need to reflect actual needs in current circumstances.

\subsubsection{LLMs to evaluate LLM}
Beyond specific metrics, ongoing research is investigating the efficacy of LLMs to evaluate LLMs (\textit{LLM-as-judge})\cite{wei2024longform, shankar2024validates}. For example, \cite{wei2024longform} et al. use an LLM to rate the factuality of a long-form response to prompts, while also using Google Search. While promising, such more complex evaluation constellations also lead to additional complexity in understanding what is being evaluated, and changes therein as the evaluator LLM changes. This leads to having to validate the validation in itself \cite{shankar2024validates}. While a human-and-LLM agent collaboration can help in this validation (as in e.g. \cite{shankar2024validates}'s EvalGen approach), the evaluation criteria cannot be fully separated from observation of model outputs, resulting in a feedback loop from output to adjusted evaluation criteria. 

\subsection{Evaluation pre/post system release}
\subsubsection{Online evaluation using actual user behavior vs. offline evaluation}
Whether evaluations are done online or offline can deeply impact results. Offline evaluations---even when using thoughtful standards---might not reflect what actual end-users do in real-life settings, or system performance over time. Online evaluations similarly are limited to which metrics have been instrumented and how actual user interactions are captured. It involves field testing; getting an IR system online and out to actual users and analyzing their interactions with the system. It can include  methods such as controlled experiments or extended A/B testing, and analysis of interactions; Hoffmann provides an overview of most common techniques used in IR settings \cite{Hofmann}.

\subsubsection{Stress testing, red-teaming and qualitative end-user evaluations}Beyond metrics and quantitative analysis oriented methods, it is crucial to apply a combination of safety/security-inspired methods, user design and UX research methods to understand the actual reactions of users. 

The logistics around red teaming can provide a good glimpse into the importance of appropriate combinations of methods. Red teaming is a common way to test LLM applications for undesirable system responses \cite{OpenAIcontentmod,OpenAIModelCard}.
Red teaming can be automated using for example sets of (generated) prompts, or done in full by human red teamers, including both the general public, or invited experts. Using LLMs as red teamers~\cite{perez2022red} by generating risky prompts at scale, or using large-scale human red teaming efforts with thousands of participants who need access points, might yield different results. Human red team approaches in which “a group of people authorized and organized to emulate a potential adversary’s attack or exploitation capabilities against an enterprise’s security posture” (if we follow the definition from the National Institute of Standards and Technology, NIST) also lead to questions about tooling, recruiting and operational process design. \citet{OpenAIcontentmod} for example provide a helpful discussion of practical data challenges in content moderation use cases. In turn, model characteristics might have consequences on red teaming results. \citet{ganguli2022red}, for instance find that RLHF (reinforcement learning from human feedback) models are increasingly difficult to red team as they scale, while they don't find similar challenges for other models. Interestingly this means that techniques such RLHF that are explicitly meant to help align agents with human preferences could also result in challenges in evaluating the systems that use them. 

This means that like any evaluation method, red teaming has to be combined with other types of stress testing, assessment of security issues, as well as evaluation of experiences of actual users. Khlaaf points out the need for carefully considering what methods and terminology are appropriate for evaluations that probe for vulnerabilities of a specific system towards the outside world \cite{khlaaf2023toward}. 

\subsection{Societal impact of a system beyond its direct implementation and use}

The impact of a system can reach much beyond its direct usage context. For example, the increasing demand for data and compute power of LLMs has environmental impact. However, such indirect impact can be hard to calculate without deep expertise. It is crucial to spend the time to evaluate evaluations methods for their suitability.
Methods have been developed in both the IR and LLM communities around reducing environmental harm\citep{scells2022reduce} and sustainability industry teams exist to ensure more energy efficient data centers for both environmental as well as monetary reasons. Others in turn try to assess whether LLMs could help in generating more green code, and develop metrics to assess the code's `green capacity' based on earlier sustainability metrics \cite{vartziotis2024learn}.

Similarly, a plethora of work points out the potential of amplifying and entrenching power structures through the usage of generative AI methods, or changing market conditions through releasing new models for free~\citep{robinsearly2024new}, de-facto changing standards to the model that gets used most in practice. However, IR and ML evaluation methods are not generally suitable for the analysis of such impact that a particular technique or system might have. Methods from political analysis and behavioral economics might be more suitable, but are generally not shared in IR or ML venues. Challenging in the evaluation of systems is a deeper understanding of the long-term incentives that are created, and the resulting ‘rational’ use of LLMs in undesirable ways. A compounding challenge is that new incentives are also necessary to ensure that interventions from actual practice can be shared. Trust \& Safety teams might be doing scenario planning or prepare for incidents and crises. 

\subsection{Sharing evaluation methods}
From the above selection of methods, which is by no means comprehensive, it is clear that practitioners have to carefully pick and choose which methods work for them. However, different organizations come from different evaluation traditions. 

Incentives to share methods and results might not align with practical product team incentives and pressures. Metrics and standards for evaluations from actual practice are often not shared in scientific literature.  Security community-style (external) red and (internal) blue teams, Trust \& Safety incident monitoring approaches, IR-communities' existing offline and online user feedback methods, or UX product testing approaches might be more (or less) top of mind depending on the organization and prior expertise. This means there is a gap in the generative IR literature in terms of shared understanding of actual practices and efficacy of methods~\cite{CramerInteractions}. If we as a community are to properly address the social risks as outlined in \ref{sec:cmr-risks-society}, it is imperative we find fast and effective ways to share these methods and align them with practical needs. Especially with the increasing speed of the field, the variety of fields involved, and volume of new techniques.

\label{tbl:methods}

\begin{table}[]
\caption{Different types of existing evaluation frameworks relevant for generative IR impact \& safety. Note this is not an exhaustive overview, but rather a quick peek at the variety of methods evaluators can (and have to) choose from}
\def\arraystretch{1.2}
\begin{tabular}{|p{.45\linewidth}|p{.45\linewidth}|} 
\hline

     \textbf{Evaluation focus }&  \textbf{Examples}\\
     \hline
    Marginal system impact, e.g. release decisions in comparison with existing technology
 &  Kapoor et al., risk framework based on cybersecurity~\cite{kapoor2024societal}  \\
 \hline
Comparison benchmarks between LLM models that include fairness, bias, toxicity-type aspects & Benchmarks used in leaderboards, e.g., HELM~\cite{StanfordHELM}, BIG-bench \cite{srivastava2023beyond}, or trustworthiness benchmarks \cite{wang2024decodingtrust}\\
\hline
 Online or offline IR metrics, including
 accuracy or quality across groups
& Online IR-evaluation methods \cite{Hoffman}, impact/fairness/bias metrics in recommendation systems \cite{smith2022recsys} \\
\hline
Evaluation metrics using automated evaluation for specific LLM techniques or risks & E.g., LLMs as agents evaluating factuality of other LLMs' statements \cite{wei2024longform, shankar2024validates} 
\\
\hline
Qualitative evaluation including human adversarial testing  & E.g., red teaming\cite{perez2022red, ganguli2022red}, and UX evaluation\\
\hline
    \end{tabular}
    
    \label{tab:methods}
\end{table}

\section{Actors, incentives and ways of getting organized}
\label{sec:incentives}

\subsection{Incentives towards misuse of AI}
Emerging AI capabilities and their consequences (good or bad) are a hot topic of discussion. But it is just as important to talk about incentives, or why individuals or organizations might choose to use AI in certain ways. 

Below are some examples of types of actors and their possible incentives that can lead to harmful uses of AI, along with ways in which some of them can be shifted in a more positive direction. AI can be transformative for human experience and quality of life, but only if incentives (both short-term and long-term) for its use are aligned with the benefits to humanity.

\textbf{Actor}: State actors and ideological groups.
\newline\textbf{Incentive}: Geopolitical influence in favor or against something. This includes the use of extra-persuasive \cite{williams2023humans}, micro-targeted content and deepfakes to sow malicious narratives \cite{simchon2024persuasive}, undermine support and trust in democratic institutions \cite{mularczyk2023row}, weaken social cohesion, etc.
\newline\textbf{Modification}: The most effective way to modify this behavior is by making it prohibitively expensive or inconvenient to use AI for these purposes, through harsh legal consequences, content moderation, or counter-speech. The burden of implementing countermeasures falls on governments, content platforms, and community organizations.

\textbf{Actor}: Criminal or unscrupulous organizations.
\newline\textbf{Incentive}: Financial gains from scams, ad-monetized website traffic, or product sales. This includes more legit-looking phishing content \cite{violino2023ai} and “Nigerian prince” letters; or gaming search engines via AI-generated SEO-friendly content \cite{orland2024lazy}.
\newline\textbf{Modification}: The incentives for financial gain are always going to exist and be exploited; protection against them can take the form of better (AI-enhanced) cybersecurity and anti-spam tools, implemented and deployed by most consumer-facing web surfaces.

\textbf{Actor}: Commercial enterprises.
\newline\textbf{Incentive}: Economic competitive advantage and increased shareholder value. Taken to its worst extreme, this incentive can lead to deceptive or discriminatory business practices, hasty deployment of cheaply developed AI to customers \cite{fowler2024turbotax}, premature restructuring of teams \cite{landymore2023sports}, etc. In the case of social media platforms, the high engagement on polarizing or sensationalist content can lead the platforms to tolerate, encourage, and algorithmically amplify it.
\newline\textbf{Modification}: The same drive for competitive advantage can also be a force for good, particularly when it is aligned with public opinion or customer sentiment. The best-case scenario is when trustworthy and safe AI makes products more usable, attracting more customers (akin to Apple’s “it just works” aesthetic that has no shortage of fans despite being more expensive than the competition). Government-led compliance requirements can also create positive incentives, like for food or car safety. And in some cases, a punitive legal strategy also works, like in the suing of tobacco companies or opiate producers, creating incentives for surviving companies to behave better.

\textbf{Actor}: Individuals.
\newline\textbf{Incentives}: Faster completion of work tasks, improved social status, revenge against perceived slights, or exploitation of the vulnerable. At worst, these can lead to cheating, misrepresentation of one’s identity of accomplishments, slander, deep-fake pornography, or AI-enhanced grooming.
\newline\textbf{Modifications}: While some of these behaviors are illegal or fundamentally antisocial (and should be prosecuted as dictated by law), the urge to improve one’s work performance or social status can be a good thing. If AI tools are designed to enhance human productivity while rewarding our creative impulses, and feel fun, joyful, and satisfying to use, people will be more likely to employ them to good ends.

\subsection{Who can shift incentives, and how}
\begin{figure}[t]
    \centering
    \includegraphics[width=0.9\columnwidth]{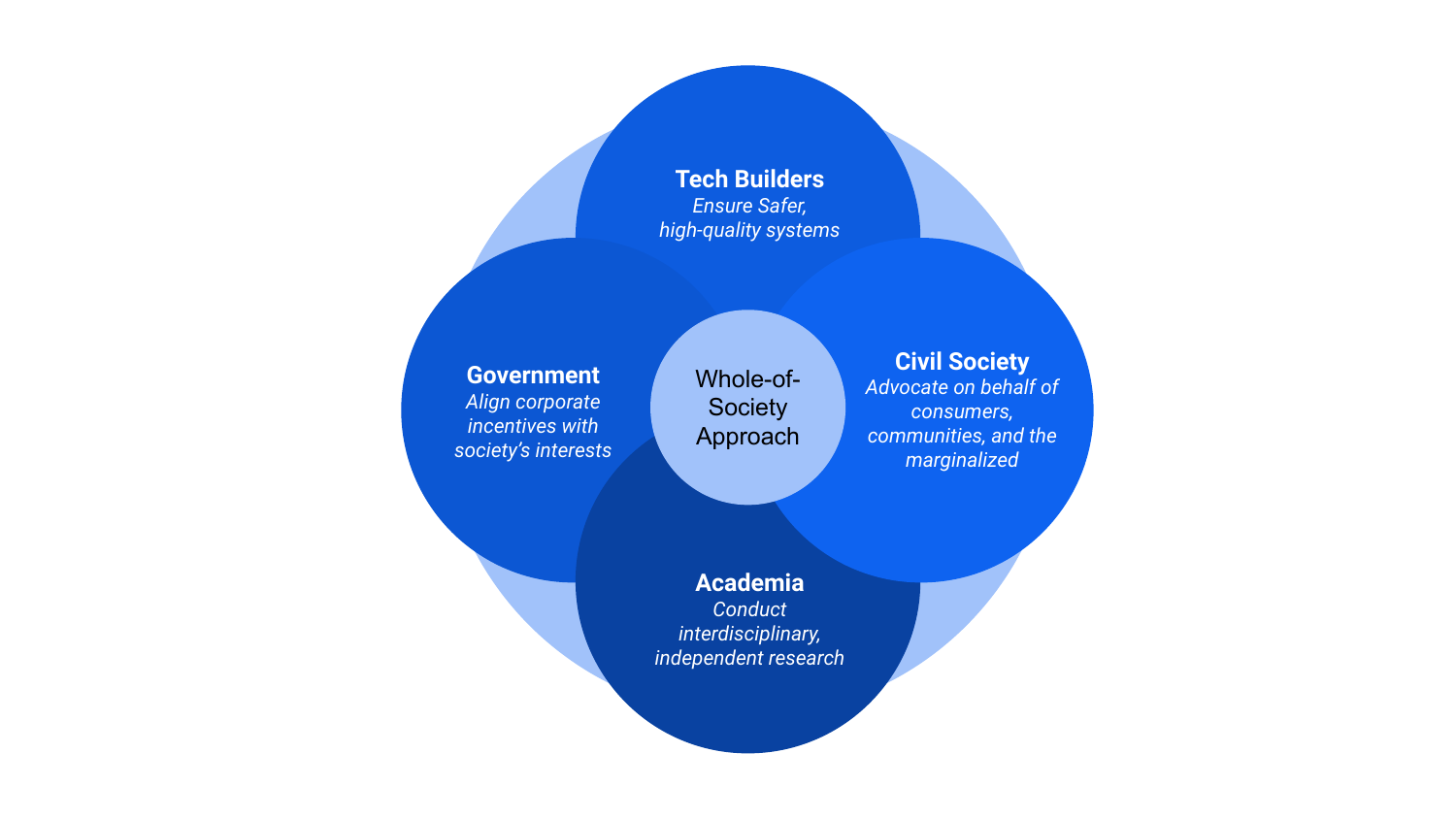}
    \caption{Primary actors responsible for aligning technology with societal interests}
    \label{fig:ai-safety-actors}
\end{figure}

In the broadest sense, it will take a whole-of-society approach to ensure that technological advances will align with the best interests of humans impacted by them (see Fig. \ref{fig:ai-safety-actors}). Technology builders (company and individual), governments, academia, and civil society all bear responsibility for ensuring that technological advances in information access align with societal interests. The rest of this section focuses on what can be done at the intersection of these groups or actors, since inter-group coordination is most often where things go awry.

\subsubsection{Organizational Factors}
While most of the literature and education in computer science by definition focuses on technical approaches, the impact of generative IR techniques can be influenced in other ways as well. 

Changing work processes \textit{within }organizations can have a direct impact on the expectations set on teams. This includes policies, explicit Go/No-Go procedures, roles and responsibilities to monitor systems, algorithmic impact assessments and model cards or other types of documentation. In different organizations, the responsibility for different measurement and mitigation might look very different. In one organization, a Machine Learning team may be expected to look at the energy consumption of their system design choices, whereas other organizations might have a technical sustainability team. In another organization, a Trust \& Safety or Integrity team might deliver evaluations of system output toxicity, whereas in another organization a separate Data Science team, or Product teams themselves, might have to do this work. In any case, if this responsibility is unclear, it is much harder to get this work done.  

\textit{External} engagement can help address internal deficiencies. Especially for audiences working on generative IR systems, some of these might not necessarily be familiar routes. 
Examples include:

\begin{itemize}
  
  \item \textit{{External advice and safety boards.} } increasingly created by companies to provide external advice for more complex safety or content moderation questions. This includes Facebook's Oversight Board \footnote{https://www.oversightboard.com}, which provides independent rulings on content moderation questions; parent company Meta's Safety Advisory Council \footnote{https://www.facebook.com/help/222332597793306/}
  ; or Spotify's Safety Advisory Board \footnote{https://newsroom.spotify.com/2022-06-13/introducing-the-spotify-safety-advisory-council/}. These do not necessarily have decision making power, but provide a more formalized way to advise external organizations and researchers.

  \item \textit{{Regulatory advisory groups and expert consultations.} } Organizations such as the UN, EU, various regions  and countries working on future AI policy have all formed advisory boards (e.g., the UN AI advisory board \footnote{https://www.un.org/en/ai-advisory-body}, the Nordic AI advisory board). Apart from such official avenues, individual lawmakers and legal firms often consult experts. While regulatory capture is a very real concern \cite{WhittakerInteractions}, this also allows for actually implementable regulation.  This means owever that considering the potential overlap between advisory boards, as well as perhaps a lack of overlap with more specific AI experts, not all relevant expertise will be represented. 

\item \textit{{Professional organizations.} } Organizations such as ACM, IEEE, AAAI, the Trust \& Safety Professional Association allow for formal and informal exchange of best practices. A major challenge is ensuring that best practices in fast moving areas are also gathered and exchanged \textit{between} organizations and to the public at large. 

\end{itemize}

For the above arrangements, getting to collections of concrete examples of what has worked in the past is increasingly important. AI developments are speeding up, and increasingly diverse professional communities are both being impacted and getting involved. This makes efficient and effective coordination even more important. For policy makers, governmental agencies and journalists it may be hard to get an overview of which professional communities can provide actionable advice---especially with new AI developments being `louder' than, for example, long-standing IR communities. 
Inside of companies, in order to benefit from external advice or research, tech teams still have to navigate how to best work with external organizations. Researchers and non-governmental organizations in turn have to know where to invest their time and expertise most effectively, and how to offer actionable advice to appropriate individuals or teams in tech companies. This includes big picture scenario planning of where to best invest, and how to create incentives that truly will have a positive impact. Implicit hierarchies of the value of different types of produced knowledge (e.g. 'being the first' or 'more technically complex'), but also a simple lack of knowledge about how certain processes work, can stand in the way of sharing of paved paths towards desired results, and of sharing these in accessible ways. It can also involve very pragmatic on-the-ground work, such as knowing how to set up contractual arrangements that work for all parties (not a skill commonly taught in IR or AI-related programs). 

\subsubsection{Data-focused methods}
While a complete overview of all different mechanisms to positively affect AI development is outside the scope of this paper, one area does provide ample inspiration. Extensive literature exists on data labor and the need to understand how to effectively advocate for that labor's value~\citep{vincent2020dont, li2023dimensions, vincent2021github, gershgorn2021github, arrieta2018should}. Especially in the realm of training data concerns, multiple practical routes already exist, including:

\begin{itemize}
    
   \item \textit{Business and partnership model development,} including developing new types of licensing and new types of business partnerships~\citep{shutterstock2023, ap2023}, along with ways to get funding to data creators. There is also also research on the efficacy of suggested mechanisms, such as data dividends that are suggested as a means of AI profit sharing~\citep{vincent2019mapping}.
   
    \item \textit{Collective action}. When new business models do not work out, coordinated action is imperative. These can be focused on data through data strikes \cite{vincent2019data}, as well as large-scale labor organizing and strikes focused on treatment of data workers. More recently the Hollywood strikes illustrated how those particularly impacted by the ways their work and likeness can be used as data, can effectively organize, lay out clear demands and succeed through both technical and organizational competence. This included understanding what incentives are at play and what leverage data producers have~\citep{vincent2023wga}. Methods include data strikes to withhold data~\citep{vincent2019data, tani2023new}, data poisoning~\citep{dickson2020What} techniques such as NightShade~\citep{heikkila2023new, shan2023prompt}, Glaze~\citep{shan2023glaze} and Mist~\citep{liang2023adversarial}.
    Ways to empower end-users and the wider public in their relationship with tech companies are important~\citep{vincent2021data}, as is understanding their potential leverage and means for protest through adjusted usage~\citep{li2019people}.
\end{itemize}

  
For effective research-informed mitigations, however, it is crucial that generative IR researchers have access to ways to learn how to effectively organize and navigate organizational and political structures, or how to communicate their results to others. Implicit hierarchies in what knowledge is appreciated in generative IR circles can become a hurdle in effectively identifying and addressing the risks outlined in earlier sections, \S\ref{sec:cmr-risks-society}, \S\ref{sec:cmr-risks-research}, and \S\ref{sec:cmr-risks-environment}. A critical factor is knowing which concrete situations matter, what to ask for in those situations and how to assess whether impacts and risks are successfully steered.

\section{Conclusion}
\label{sec:conclusion}

\begin{figure}[t]
    \centering
    \includegraphics[width=0.7\columnwidth]{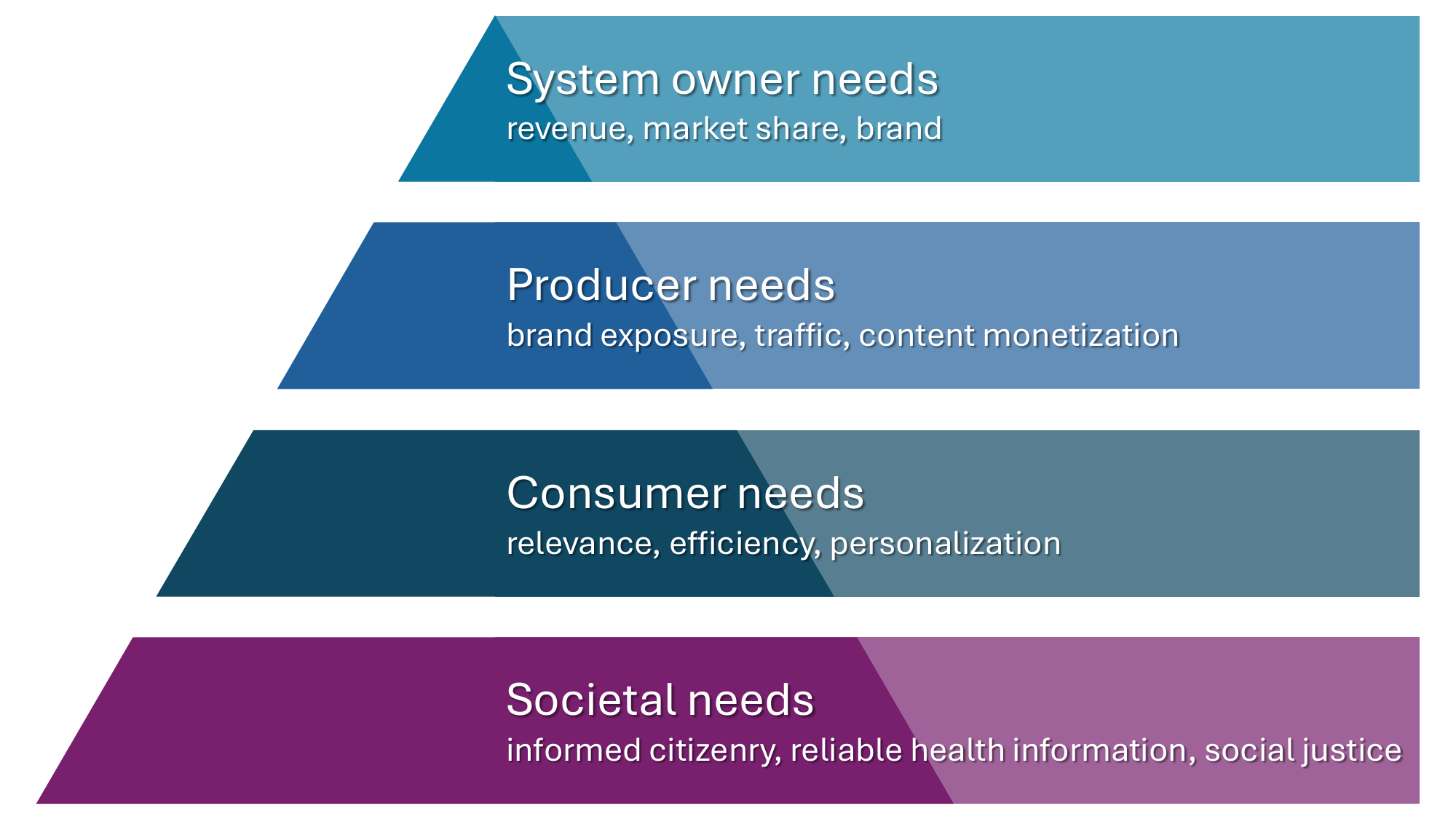}
    \caption{Mitra's \citep{mitra2024search} hierarchy of IR stakeholder needs.
    More critical needs are at the bottom of the pyramid.
    This figure has been reproduced from the original paper with permission.}
    \label{fig:hierarchy-of-stakeholder-needs}
\end{figure}

In this chapter we have presented a discussion on the sociotechnical implications of generative AI for information access.
These deliberations are grounded in how these emerging technologies are currently being applied in IR applications as well as their future applications as being envisioned by practitioners and researchers.
It is important to recognize that sociotechnical visions of what information access should look like in the future are not just shaped by what emerging technologies like generative AI make plausible, but  that visions for the future of information access in turn shape AI technologies themselves.
\citet{mitra2024search} proposed the hierarchy of IR stakeholder needs shown in \ref{fig:hierarchy-of-stakeholder-needs} and argued that IR research and system development require a fundamental shift towards re-centering societal needs and that we should reimagine information access as a vehicle for alternative futures.  
When contemplating the implications of emerging technologies, we risk of falling in the trap of limiting ourselves to how the technology (and its process of development) is today, rather than how it can be or \emph{should be} in the future.
Neither generative AI nor its application in the context of information access is predetermined.
So, while it is important that we consider potential harms of contemporary applications of generative AI in the context of information access, we close with some open question for the reader:
\emph{
If not this status quo, then what---and especially how?
What is the future of information access that we want to imagine for our collective wellbeing, and how can generative AI be another tool in the toolbox towards that transformation?
}



\bibliographystyle{abbrvnat}
\bibliography{references}
\end{document}